# TAXATION AND SOCIAL JUSTICE

Prof. Boyan Durankev, PhD, UNWE

## Introduction

The link between taxation and justice is a classic debate issue, while also being very relevant at a time of changing environmental factors and conditions of the social and economic system.

**Technologically** speaking, there are three types of taxes: **progressive, proportional and regressive**.

From a **practical** standpoint, taxes have been around for as long as there have been countries. According to the Bible, a "tithe" of the harvest had to be set aside and given to the priests. Later, as late as 1913 in the USA, a constitutional amendment was adopted giving the federal government the power to levy an income tax.

## 1. Theoretical foundations of modern taxation

The classical political economy, as well as the economics later on, deal with the income tax within the context of a comprehensive **tax system**. This "classical" outlook is not contemplative, but of a deeply social and justice-seeking nature.

In 1776, in his most famous work "The Wealth of Nations"[1] **Adam Smith** discussed at length the taxes upon the wages of labour. The classical economist strongly believed that the money price of labour is regulated by two circumstances: the demand for labour (whether it requires an increasing, stationary, or declining population), and the average price (and its fluctuations) of provisions. Consequently, if the two factors remain the same, then "*a direct tax upon the wages of labour* can have no other effect than to raise them somewhat higher than the tax." The view of social justice is implied: "The persons, besides, who enjoy public offices, *especially the more lucrative*, are in all countries the objects of general envy, and a tax upon their emoluments, even though it should be somewhat *higher* than upon any other sort of revenue, *is always a very popular tax*." In other words, direct taxes, as well as indirect ones, *should not eat up most of one's earned income*, but *higher income should be taxed at a progressive tax rate,* which, from a justice point of view, is "always a very popular tax" for society.

---

[1] *Smith, Adam*, The Wealth of Nations. Partizdat, S., 1983 (1776), pp. 841-844. (in Bulgarian)

In 1817, **David Ricardo**, who had a thorough knowledge and understanding of the works by Adam Smith (whom he severely criticized for some of his theoretical considerations about taxation and tax) and Thomas Malthus, focused most closely on taxation and especially on the taxes upon the wages of labour[2]. Ricardo took a step forward – not only in terms of tax collection, but also in terms of tax revenue use. He wrote that "*the fund raised by the tax is employed by Government in maintaining labourers*". In addition, "when taxes operate *justly*, they … raise from the people as little as possible beyond what enters into the public treasury of the State"; "Dr. Smith uniformly, and I think justly, contends, that *the labouring classes cannot materially contribute to the burdens of the State*". He then went on to establish what "the *just share of the taxes*" was.

Much later, **Karl Marx** included "*labour power*" (not "labour"!) and the direct taxes a worker pays in *the price of commodities*[3]. According to him, the Capitalist tax policy was "*unjust*" due to the fact that the State, being a representative of the bourgeoisie, sought to place the tax burden on the working class.

More recently, several hundred elite economists have focused their research efforts on **income** and **income tax systems**. The paper does not need to present a complete literature review. Just a few general theses and hypotheses pertaining to the last decades are going to be highlighted.

In theory, the "income" concept seems simple, but in practice it is extremely difficult to define **what is and isn't "income"**. The State runs into additional difficulties by taxing *different types of income at different tax rates* and providing *different types of tax reliefs*. Any difference in terms of income, rates or reliefs introduced by way of tax laws requires a set of precise legal definitions, and when the legal definitions do not fit well with the logical economic ones, taxpayers are clearly urged to earn their income in such a way as to have it taxed at a lower tax rate.

There is a difference between **tax evasion** and **tax avoidance**, although it isn't always clear-cut. In EU member states, there is a flourishing *market for activities and services related to tax avoidance*. The fact that on the whole there is a great interest in tax loopholes by exploiting different tax reliefs, and the broad publicity given to such cases, give the public the impression that tax laws are drafted with specific individuals in mind and deliberately create *tax injustice*.

---

[2] *Ricardo, David*, On the Principles of Political Economy and Taxation. Partizdat, S., 1981 (1817), pp. 205-229. (in Bulgarian)

[3] Marx, K, Engels, Fr. Works. V. 23. Publishing house of BCP, S., 1968. (in Bulgarian)

The reasons for **introducing some of the criteria** for different tax reliefs are very often meaningful and justified. For instance, the possibility to deduct medical expenses from the pretax income available in a number of countries was introduced due to the belief that the patients, who pay exorbitant amounts of money for treatment, should be treated differently from able-bodied people. Education expenses are also treated in the same way, especially in light of the fact that higher education acts as an "economic escalator" moving people up from poverty to wealth, and also taking into account the fact that the more educated a society is, the more productive it is compared to less educated ones. Also, a large family with children has the objective necessity to spend more money than another family (without children) earning the same income. But no matter how legitimate the justification seems to be, the consequences – namely, complicated tax laws and increased possibilities for tax avoidance – are a matter of general concern.

Apart from this, these special conditions affect not only **the size and structure of the incentives**, but also **the subsequent costs** associated with these incentives. The possibility to deduct medical and educational expenses or childcare expenses (which in reality constitute an investment) may tempt people to spend more on expensive drugs, multiple educational programmes or goods intended for adults (and not ones directly related to childcare) than they could otherwise afford.

That's exactly why there are numerous "tax shields" available worldwide. These are expenses and investment schemes related to tax incentives that could help people *to reduce their tax liability* and have gained immense popularity and enjoyed increasing demand. These schemes undoubtedly allow some people to increase their after-tax income, and they also turn into an excellent source of income for those tax consultants and financial advisors that apply them most successfully. Not all taxable persons have equal (financial and legal) resources in order to gain access to these "tax shields"; the latter provide greater capital allowance reliefs, i.e. on interest, capital gains, etc. rather than on salaries. Individuals who rely mostly on their salaries and other income, such as fees for services rendered, have considerably fewer opportunities for tax avoidance; or are convinced that they have fewer such opportunities. Actually, a big part of their extra service pay allows tax avoidance even for this form of remuneration.

A vicious cycle is thus created: **due to the numerous incentives and special tax reliefs, tax rates become higher than they would be if there were equal treatment in order to collect the necessary revenue for the state budget.** Higher rates in turn increase the demand for "tax shields" in order for tax payers to become eligible for special relief which additionally reduces the tax base and, in turn, results in the **further increase in tax rates** for subsequent periods. The snake eats its own tail.

Higher taxes and tax rates lead to a decline in economic efficiency and to **lower rates of savings and return**, thus **reducing the opportunities** for higher consumption and higher GDP (related to the higher consumption), and the increase in economic activity leads to technological innovation. However, the public (with the exception of those benefiting from "tax incentives" and different reliefs) is upset by the blatant *injustice* caused by the existing tax system and its incentives.

Why is it so difficult to design a **fair and effective tax system**? Aren't there any simple principles and clear criteria that can be used to evaluate the alternative tax systems? There are several main principles, but as they are more than one, some *compromises* have to be sought among them. People's opinions on the relative weight that should be attributed to each criterion differ.

It is now widely acknowledged that there are **five desirable principles for a "good" tax system[4]**:

1. *Economic efficiency*: the tax system should not interfere with the efficient allocation of resources.

2. *Administrative simplicity*: the tax system should be easy and relatively inexpensive to administer.

3. *Flexibility*: the tax system should be able to respond quickly (in some cases automatically) to changed economic circumstances.

4. *Political responsibility*: the tax system should be designed so that individuals can ascertain what they are paying and evaluate how accurately the system reflects their preferences.

5. *Fairness*: the tax system should be fair in its relative treatment of different individuals.

All of the principles mentioned above affect both **economic growth** and **the behaviour of individuals and families**. Therefore, the type of tax system should be determined by applying the **systematic approach** that will eliminate the possibilities for both economic and social development. *No universal tax system exists* that can be transposed directly in Bulgaria, for instance. Either the economic factors or the factors of social justice and solidarity will dominate at one point or another.

---

[4] *Stiglitz, Joseph*, Economics of the Public Sector. University Publishing House "Stopanstvo", S., 1996 (1988), pp. 454-455. (in Bulgarian)

**Economic efficiency** is an extremely important and, in most cases, fundamental principle. If there is a wide range of tax reliefs and incentives, individuals and families can improve their financial situation by applying successful tax avoidance efforts instead of directing these efforts towards elaborating good business projects or producing more and higher quality work.

A tax is *undistorting* only if the person is unable to do anything to change his/her tax liability. **Taxes that are not distorting** are called by economists "**flat-rate taxes**". *Distortions* occur when individuals attempt to reduce their tax liabilities. All taxes are virtually distorting. The poll tax, which was owed by everybody regardless of their income or wealth, is a flat-rate tax; a tax that is based on unchangeable characteristics (age, gender) is also flat-rate.

Taxes such as the tax on wages and on the return on capital change the **economic equilibrium**. The tax on savings account interest may decrease the amount of the savings and eventually reduce capital stock, which, in turn, causes a drop in workers' productivity and a fall in their wages. These indirect tax effects are called effects on the general equilibrium. The effects on the general equilibrium have important distributional implications which sometimes move in a direction opposite to what legislators have intended.

## 2. Justice and taxation

No tax system in the world can be regarded as **completely fair**. Citizens around the world describe their countries' tax systems as **unfair, but from a different standpoint.** It is difficult, however, to determine what exactly is fair and what isn't in a specific place and at an exact moment in time.

The **concept** of "**justice**" itself is very often defined by resorting to artificial constructs rather than to a phenomenon (process) of universal nature[5]. And if there is a *generally accepted* form of justice somewhere, then it is not universal or universally applicable[6]. Every society is just in its own way, but it is generally acknowledged that nowadays *social justice is undergoing a process of development.*[7]

---

[5] *Rawls, J.*, Theory of Justice. Sofia S.A. Publishing House, S., 1998 (1991). (in Bulgarian)
[6] *Walzer, Michael*, Spheres of justice. Critique and Humanism Publishing House, S., 2010 (1983). (in Bulgarian)
[7] *Huddleston, J.*, Search for a just society. University Publishing House "St. Kliment Ohridski", S., 2001 (1989). (in Bulgarian)

But what is the **modern understanding of justice**? There is no justice in the laws of nature or even in the Universe; throughout human history, justice has also been more of an imaginary concept, *an imagined reality* than a law of nature. [8]

**Freedom** is what gives room for personal development; **justice** is what creates a large-scale cooperation network among free individuals; **solidarity** is what provides an additional social network in order to further this cooperation and reach the weaker members of society.

For example**, in a decentralised capitalist market economy freedom** creates a lean hierarchical system based on the results of the activity of the "invisible hand". Thanks to this invisible hand, some become very rich, while others remain very poor; some enjoy privileges and power, while others, who are lower down the hierarchical ladder, remain repressed and disempowered; in a capitalist market economy *inequality is fair*; transferring the inheritance of the family to the children is also fair; equality before the law is also perceived as supreme justice. Such a vertically expanding society, especially in the context of a developing democracy, however, gives rise to **institutions and forms of justice that are a far cry from the principles of freedom**: free education for all, health care for all, unemployment benefits, etc. The idea that freedom is unlimited is a *fiction*; the idea that justice requires (in the 21$^{st}$ century) helping the poorest and weakest is also a *fiction*. But in all these cases, justice – as perceived by the greater part of society – plays the **unifying role of social glue**.

Another fiction upheld by the rich is that they have more money and benefits because they are more capable, intelligent and enterprising. *They* believe that it is only **fair** that they should eat better food, wear more expensive clothes, and provide the best education for their children. *They* believe that it is **only fair** that homeless, landless, poor and hungry people should exist – they are just lazier, less resourceful and more stupid. This is as fair *for them now* as it was until recently for people of colour and women to have fewer rights. All **these kinds of justice are a figment of the human imagination, but without them society cannot be internally unified** and will fall apart.

*The present Western* **freedom and justice** and the social hierarchy, mythology and fiction they have created are considered to be natural and universal, whereas all "non-Western" types of freedom and justice are considered to be false and ridiculous. In most Western societies, **Western justice** requires wealthy families to have even wealthier children, provided with better education, health care and living conditions simply because they have been born in wealthy families; most poor children will *justly* remain poor all their lives because they have been born

---

[8] *Harari, Yuva, Noah,* Sapiens: A Brief History of Humankind. East-West Publishing House, S., 2016 (2011), p. 121. (in Bulgarian)

in poor families. Of course, some people are naturally more gifted than others; but in a Western-style society, two children with the same abilities will develop differently if one of them comes from a wealthier background.

All sophisticated human societies have put in place imagined realities, hierarchies, freedom and justice that incorporate discrimination, including based on income. Not all hierarchies are morally identical and equally discriminatory. In modern societies, **the evolution of freedom (and its related "market" justice)** has a *sharpening* effect and stretches the hierarchy, whereas **the evolution of democracy (and its related "social" justice[9])** doesn't have a sharpening, but a *rounding effect* on society. One does not negate the other. The evolution of freedom can happen in parallel to **the evolution of justice**.

In societies where more emphasis is put on **market justice** („every man for himself") *the rich will "naturally" pay lower taxes* than in other analogue societies. In such societies the hierarchical structure will be much more stretched and sharpened. In societies where more emphasis is put on **social justice** ("all for one and one for all") *the rich will "naturally" pay higher taxes* in order to make the structure flatter and more rounded and provide *equality of access*.

There are two main **concepts of tax justice**: horizontal equity and vertical equity.

A tax system is considered **horizontally equitable** if the people who are the same in all relevant respects receive the same treatment. The principle of horizontal equity is so important that it has been incorporated in all Western constitutions. Thus, a tax system that contains gender, race, skin colour or religious biases is considered to be horizontally inequitable (and unconstitutional)[10].

Although the basic idea is quite straightforward, there are some ambiguities as regards the definition. What does two people being the *same* in all relevant respects mean? And what does two people receiving the same *treatment* mean?

Let's look at two people who are the same in all respects except the fact that one of them has bought himself an expensive watch, and the other – a cheap one. Does the tax system treat both of them in a horizontally equitable way if it taxes the two different watches at different rates – the former is treated as a luxury item, and the latter is treated as an ordinary consumer product? One of them pays *more taxes* than the other (for the same good that even fulfils the same function) and in this context the tax system seems unfair. But both have the same "set of capabilities". The man who bought an expensive watch could have bought a cheaper one if he wanted to (and vice versa). The tax system does not discriminate; it does not differentiate between people. In this example we have two goods that "essentially" fulfil the same function. In practice, there are numerous examples where the tax system treats people with different tastes differently – higher taxes on concentrated alcoholic beverages discriminate against people who prefer Scotch whiskey to French wine or Bulgarian beer. People who prefer to spend their holidays in their own villas receive preferential treatment compared to those who prefer to stay in a hotel.

---

[9] Justice is a fictional concept invented by people. Therefore, the adjective "social" is meaningless, but sounds good.

[10] *Stiglitz, Joseph*, Economics of the Public Sector. University Publishing House "Stopanstvo", S., 1996 (1988), p. 465. (in Bulgarian)

If we assume that *the differences in taste are significant economic differences that can be taken into account by the tax system*, then we can say that the principle of horizontal equity does not apply here. The two people in the example above are not the same in all relevant respects. By going to such extremes, the principle could quickly be rendered meaningless: *no two people are exactly the same*. What are **the acceptable differences**? Sadly, the principle of horizontal equity does not provide an answer to this question.

The first assumption may be that *all differences are unacceptable*: gender, age and marital status must be relevant. Now, in practice, a distinction is made by age (people of retiring age receive additional tax relief because the pension is not taxed and constitutes personal income) and by marital status (in countries where there is a family income tax). Apparently, the legislator has deemed these differences to be relevant.

Perhaps *age and marital status are relevant* because they affect the ability of individuals to pay. But if this is an acceptable basis for differentiation, are there also other acceptable bases? For example, whether the fluctuations in economic costs associated with the taxation of different groups are a legitimate basis for differentiation? It has been proven that tax-induced inefficiency depends on the magnitude of the responses (sensitivity) to taxes. In a household with two wage-earners, the one earning the lower wage (in many cases the woman/mother who has fewer opportunities for career growth) shows a much higher wage sensitivity than the principal wage-earner of the household. If the state were concerned about reducing the inequity caused by the tax system, it would tax the woman/mother of the household at a lower tax rate. But is this fair?

The following example will show *how difficult it is to even determine the meaning of fair treatment*. Let's assume that we agree that a man and a woman who have received the same income throughout their working lives should be treated equally for social security purposes. Should the total amount of their pensions be the same for both the man and the woman, or should the annual amount of their pensions be the same? On average, women live much longer than men, so the two approaches will produce different results. If women receive the same annual pension as men, then the total expected amount of their pensions will be much higher than that of men. A lot of people find this unfair.

In other words, **horizontal equity is often nothing more than an "imagined" reality**.

While the principle of horizontal equity states that people who are essentially the same should be treated equally, **the principle of vertical equity** stipulates that **some people are able to pay higher taxes than others**, and are obliged to do so. With this principle we run into three main problems: to determine *who*, in principle, should be taxed at a higher rate; to decide *how much more* someone should be taxed than others if they are able to pay a higher rate; and to apply this into practice, i.e. *to elaborate tax rules* in line with this principle.

Usually, there are three criteria for the assessment of **whether an individual needs to pay more than someone else**. Some people may be deemed to be able to pay more; others may be deemed to have a higher level of economic welfare (for example, larger homes or annuities); yet others may be deemed to receive more benefits than the total government expenditure.

Even if a consensus is reached on which of the three criteria to be applied, there will inevitably be disagreement on **how to measure one's ability to pay** – based on their economic welfare or the benefits they have received. In some cases, the same indicators, such as income or consumption, can be used to measure both one's ability to pay and one's economic welfare.

However, most democratic societies have decided that richer people are able to pay higher taxes, and are obliged to do so. Vertical equity, even though it is an imagined reality, is more widely accepted. *"Greater equality" is seen as fairer than the growing inequality*.

**There is no unanimity regarding the choice of a suitable tax base**.

Firstly, let's analyze the idea that **wealthier individuals should contribute more**. The critical question is *how to determine whether one person is more affluent than another*. Let's look at an example of two different individuals. *One* is proverbially hard-working, works 10-12 hours a day, 6 days a week. He has no free time, will delay marriage, will spend little time with his future children; but he has inherited a large home, a beautiful villa and a new sports car as well as substantial savings in the bank. *The other* is a notoriously lazy and happy-go-lucky guy. He works 8 hours a day, 5 days a week in order to earn some income. He parties hard, gets married more than once, has several children and looks after them reluctantly. Certainly, the consensus will be that the latter is in a better position than the former, but *the tax system will tax the former at a higher rate than the latter*. **The tax system must obviously be based on a narrowly defined concept of welfare. It should not measure total, overall**

**welfare, and thus it is inherently inequitable.** This may, for example, be juxtaposed with the division of duties and benefits within the family. In this case, it is possible to make a thorough assessment of both needs and capabilities. We may spend more money on a child who is born at a later time when the family earns higher income. This is fairer. It provides much fuller information than the state can ever get.

Secondly, let's look at an example of individuals who are equally wealthy at the start of their careers. The differences don't lie in their starting wealth, but in the income they earn. According to this system, **people earning higher income will pay higher taxes than those earning a lower income**. Let's imagine, for example, a pair of twins who have been born identical and have equal opportunities for development. One of them loves nuclear physics, studies at university for a long time, undertakes doctoral studies, and gets a low-paying job as a scientist at a research institute. He is quite content, but his salary is low. The other one studies for a shorter period of time, but becomes an alcoholic beverage entrepreneur and earns a high income. However, he is not happy. *Their economic capabilities, what they could do, are the same*. They knew they had the same opportunities to make money. Yet *their choices differ*. One of them has a low income, and the other – much higher. Is it fair for the latter to pay much higher taxes than the former? A lot of societies agree that economic capabilities do not serve as a fair tax basis. What they deem to constitute a fair tax basis is *the extent to which individuals have taken advantage of the opportunities that society has offered them. The real income of the former twin is smaller than the potential income that he would have had if he had pursued a more lucrative profession.* However, society has decided that **the real income should serve as a suitable tax basis.** Yet, a lot of people think that the potential income and not the real one should be the relevant indicator. The example above illustrates two points: firstly, the two "identical" individuals differ in terms of whether the economic welfare or the ability to pay should be accepted as a suitable tax basis, similar criteria (i.e. income, potential income, consumption) will be used in practice; secondly, even if both agree that the ability to pay is a suitable tax basis, the controversy of whether this ability is to be measured in terms of real or potential income will remain. It is virtually impossible for the tax system to be based on people's potential capabilities. In most countries, *the real income* is used as a basis for measuring economic welfare. But in other countries, the pay rate is supposed to be a better indicator of personal economic capabilities than real income, as the income tax makes people who choose to work more pay more.

However, it is also widely believed that neither of the two constitutes a "fair" tax basis. Both indicators mentioned above correspond to a individual's contribution to society, and to the value of his/her economic "output". **Isn't it fair r for people to be taxed on the basis of what they purchase** (by way of an indirect tax, e.g. VAT) rather than on the basis of what they earn – in other words, on the basis of their consumption, and not on the basis of their income?

The following example once again illustrates **the polarised views on justice**. The difference between income and consumption is saving, i.e. the income is either spent in full or some amount of it is saved. So the main problem is whether savings should be exempted from taxation, leaving only the tax on spending.

Let's look at another example of two people who get the same salaries. The first saves 30% of his income a year, and the second spends everything he receives. Over the years, throughout their working lives, they pay different indirect taxes, of which the latter pays more. Upon retirement, they receive the same pensions, but the former has accumulated substantial savings. The latter receives social assistance because he doesn't have any savings. The former, however, pays considerably higher taxes than the latter because he is taxed on his savings interest, and does not receive any benefits from the state. The former naturally thinks that *the tax system is unfair*, as both men have the same set of economic capabilities. He will ask himself whether the state should force him to pay taxes (on savings) when there is delayed consumption and whether the state should assist someone who did not think about the years after retirement. "Why am I being punished with additional taxation, while someone else, who has lived a lavish lifestyle, is being rewarded?" The latter will say that the past does not matter, as their incomes are different upon retirement.

Nowadays, the position of the saver enjoys much greater support than the one of the spender, partly because of the problem of justice, and partly because of the need to make provisions for the future. According to the view that rallies more and more supporters, the lifetime, hourly, daily or even annual incomes constitute appropriate tax bases. The lifetime income is defined as **the present discounted value of personal income**.

## 3. Income tax and social justice in the modern world

Although the principle of "equality of access", which is axiomatically linked to reducing inequality, underlies most practices in European countries and the USA, **even with progressive taxation, the share of income earned by the top 1% of the population for the period 1975-2014 continues to grow**. For example, in the USA, the top 1% increased its share from 8% in 1970 to 17% in 2010. Moreover, in the USA there is a fully-fledged tax credit system[11], including child care expenses[12]. Direct taxes, calculated as a percentage of the federal revenue from various sources, increased from 16.9% in 1940 to 44.0% in 1960, 47.2% in 1980 and 40.3% 2016. [13]

---

[11] https://www.irs.gov/uac/about-form-1040

Table 1

Share of the income earned by the top 1% for the period 1975-2014[14]

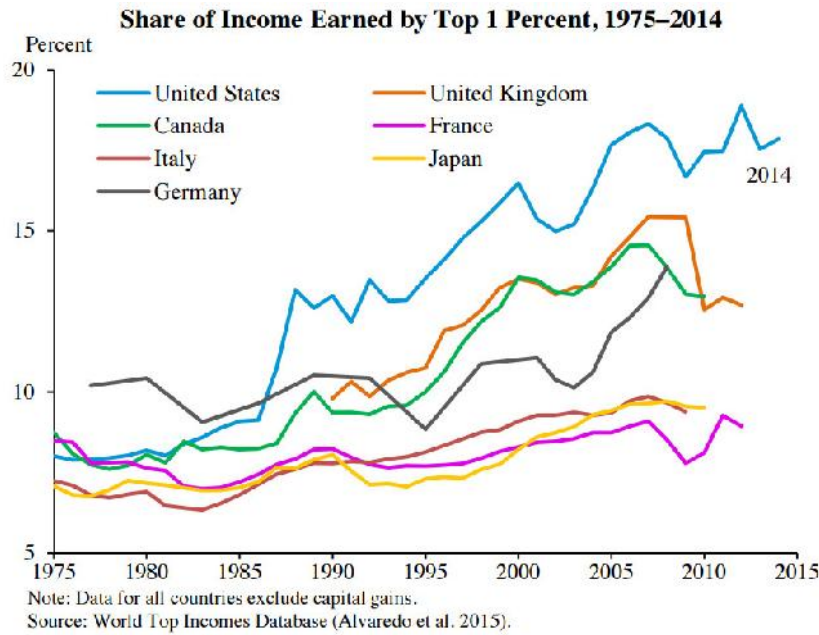

## 4. Income tax and social justice in Bulgaria

After the Second World War, and especially after the nationalization of private property and the transfer of power to the Communist Party, the whole economic activity in the People's Republic of Bulgaria operated under **the command economy system**. The efforts of the socialist state were focused on d**eveloping and implementing the annual and five-year national economic plans**. The plans themselves were based on a system of national economic balances: material, labour and financial resources.

Although the financial resources were an extremely important tool in the hands of the socialist state, they were of somewhat **subordinate importance**. The implementation of the national economic plan required the mobilisation of the state budget, people's taxes, long-term and short-term credit, insurance, savings, and the financial resources of socialist enterprises and organisations.

---

[12] http://www.taxpolicycenter.org/briefing-book/what-child-tax-credit-ctc
[13] https://www.census.gov/content/dam/Census/library/publications/2016/econ/g16-qtax2.pdf
[14] Economic Report of the President. February 2016. https://www.whitehouse.gov/sites/default/files/docs/ERP_2016_Book_Complete%20JA.pdf, . 25.

During the first period between 1944 and 1949 the tax system was simplified by adopting the Turnover Tax Act and Income Tax Act. Under the second law, **the progressive tax rate reached 84%** of the income earned.

In the 1960s, and especially the 1970s, the tax system was refined. The income tax (the tax on wages and similar remuneration) accounted for the smallest share of government revenue. They were taxed at the lowest rates. **An income tax threshold** was established – the minimum wage of BGN 120. The premiums given to champions and winners in competitions, marriage grants, childbirth grants, assistance in the case of illnesses and accidents, child allowances **were tax exempt**. The progressive tax rates which were applied ranged from 8 per cent for income from BGN 120 to BGN 130 to 14 per cent for a monthly salary of over BGN 340. In the mid-1980s, salaries reached the maximum of the salary range and the tax virtually became **a "covertly" proportional tax**. The gross annual fee income of scientists, artists and cultural workers was taxed separately at an unevenly increasing progressive rate reaching up to 50 per cent for an annual income of over BGN 40,000. This tax was definitely imperfect as it was charged on the gross income, and not on the net one. Income from rents, annuities, dividends, and other sources were taxed at the highest rate, because the socialist system treated these incomes as non-labour. The income from collective and personal work put into the production and sale of agricultural products, as well as the income of engineers and technicians earned outside the workplace were taxed at **a moderately progressive rate**. After 1987 some incomes became subject to a patent tax.

**The rental income** was taxed at rates ranging from a minimum of 9% to a maximum of 81%. People who did not have a family with children also paid the so-called "bachelor" tax of 10% on the income. The social security contributions were 30%. On average, the corporate income tax reached as high as 62%.

Under the planned economy regime, the **taxes and fees collected from the population had an extremely small relative share of the total amount of budget revenues**. In 1958-1960 the total budget revenues amounted to BGN 7,838 million, whereas the taxes collected from the population were in the amount of BGN 486 million. In 1980, the total revenues amounted to BGN 13,098.6 million, whereas the taxes and fees collected from the population were in the amount of 1,147.8. In 1987, the total revenues amounted to 21,396.8, whereas the taxes and fees collected from the population were in the amount of BGN 1,721.3 million. Due to the strict financial control **tax collection efficiency was almost 99%**.

At the beginning of 2008, the legislators decided to turn the personal income tax into the well-known "**flat tax rate**" of 10 % for all Bulgarian citizens regardless of the size of their income. Indeed, under pressure from IMF and the World Bank the flat tax rate was adopted not only in Bulgaria, but also in several other new EU member states (Latvia, Lithuania, Estonia, Romania and the Czech Republic), where it was set at a significantly higher percentage. The flat tax rate did not receive a warm welcome from everyone[15].

Here are the **main repercussions** of its adoption:

1. the top 2-3% of the population are left with much more income, which results in an economic restructuring directed at the production of luxury goods.

2. society is divided into a large number of poor people and a small number of super-rich people.

3. traditional values such as "solidarity" and "justice" are shifted towards greater market freedom.

Last but not least: it does not *automatically* decrease the shadow economy and does not automatically attract foreign investment.

**The disadvantages** of the flat tax rate:

1. Slowdown in economic growth (factually proven by Joseph Stiglitz, Paul Krugman, and Thomas Piketty).

2. Rising income polarisation and class division into (1) super-rich and (2) super-poor.

3. Opportunities to "buy" elections.

Calculations have shown that the flat tax rate affects low-income earners the most. For example, those people whose salaries are around the minimum wage – BGN 460 since the beginning of this year – will have their income automatically slashed by BGN 46. And here comes the legitimate question about the so-called income tax threshold.

In Bulgaria "the flat tax rate" *is not exactly* flat. **The tax reliefs** provided for by law in 2016 are as follows:

- for additional voluntary contributions. This tax relief is in the amount of up to 10% of the amount of the annual tax bases for personal contributions for additional voluntary social security paid during the year

- for voluntary health insurance and life insurance. The tax relief includes both types of insurance and is the same as the above.

- for personal contributions for insurable length of service upon retirement. This tax relief applies to people who are of retiring age but need 5 more years of insurable service in order to receive a pension. In this case, they are entitled to purchase insurable length of service

---

[15] Zhuleva, N., The sin of the political class against Bulgaria. Duma newspaper, 13.11.2009, issue 259. (in Bulgarian)

and the tax base is reduced by the amount of the social security contributions paid for this purpose during the year.

- for donations. Up to 5% of the amount of the annual tax base when the donation is made to certain institutions.

- for young families to buy a home on credit. The amount of the tax base may be reduced with the interest payments made during the year. If the loan is bigger than BGN 100,000, the tax relief applies to interest payments made on the first BGN 100,000 of the principal of the mortgage credit.

- for children. The tax base is reduced by BGN 200 for one child, by BGN 400 for two children and by BGN 600 for three or more children. For children who have a disability rating of 50% or more, the tax relief is BGN 2,000.

- for people with reduced working capacity. The tax base is reduced by BGN 7,920 for people whose working capacity is reduced by at least 50%.

## 5. Opportunities for progressive taxation and increased social justice in Bulgaria

For EU citizens and Bulgarian citizens in particular, **tax rates and payments** are particularly sensitive issues. On the one hand, the idea that it is only fair that the richest (companies and citizens) should pay more is almost consensually accepted[16]. The fact that the so-called top 1% are super-rich is well-known and the other 99% are willing to support a higher tax on them[17]. On the other hand, the covert transfer of personal income and personal consumption into "corporate accounts" is a trend which society frowns upon. On top of that, corporate accounts are often hidden in tax havens. On yet another hand, there is a widespread belief (especially in Bulgaria) that the tax system is corrupt.

There are many studies that compellingly prove that **the adoption of the flat tax rate in Bulgaria has resulted neither** in greater economic growth, nor in a more rapid inflow of foreign capital, nor in a higher level of social justice[18].

---

[16] The national representative survey "Attitudes to the tax policy in Bulgaria" conducted at the end of 2016 shows that more than 50% of respondents believe that the tax system exacerbates economic inequality; 22.7% believe that the tax system does not affect income distribution.

[17] *Stiglitz, Joseph*, The price of inequality. How today's divided society endangers our future. East-West Publishing House, S., 2014 (2012). (in Bulgarian)

[18] *Ganchev, Gancho*, Flat tax rates or flat delusions. http://bg.mondediplo.com/article174.html; *Angelov, Ivan,* The truth about the low "flat" tax rates. http://epicenter.bg/article/Istinata-za-niskite--ploski--danatsi/115218/11/33 (in Bulgarian)

Justice requires **changes in all three directions**: higher taxes for the richer taxpayers, better collectability of tax payments and uncorrupted use of collected taxes. Research shows that the greater the equality is and the flatter the society is, the stronger and more stable they are[19]. Estonia, Slovakia and Ukraine have abolished the flat tax rate.

**But what are the possible options for change?**

Firstly, there is the principle that the tax system is indeed a *system*. One tax change affects all others.

Secondly, a number of *state policies* are implemented through the tax system. If, of course, such policies exist.

Thirdly, the statement that "tax cuts are easy, but tax increases are difficult" is true.

**What is to be done?**

1. The information necessary to **determine the manner in which the balance between personal income and corporate income taxes should be changed** can be obtained only from the system of national accounts. The Ministry of Finance, not NSI, has the necessary information.

2. **The ratio of direct taxes to indirect taxes** does not have to change if there is steep income growth. Such a steep growth cannot be expected, so indirect tax differentiation can also be considered. As practice has shown, the abolition of or the reduction in VAT on certain groups of goods (bread, textbooks, etc.) does not automatically lead to a reduction in their prices but changes the payroll-to-gross profit ratio in favour of the latter. It is the manufacturer, and not the end-user, who gains.

3. **The future state policies** are not clear, at this moment anyway.

3.1. Let's look at one example: preventing the demographic catastrophe. All forecasts indicate that the population of the Republic of Bulgaria is on a catastrophically declining path[20]. In 2015, the population was 7,168,009, and according to a realistic projection in 2030 it will be 6,554,784, in 2050 – 5,813,550, and in 2070 – 5,132,023. In economic terms, if Bulgaria wants to "produce" more children who can live in better conditions, **the adoption of a household income tax** is a real objective necessity. Conversely, the man and the woman will not be taxed separately, but there will be a tax on the income of 2+N (the man, woman and the number of children).

---

[19] *Wilkinson, Richard and Kate Pickett*, The spirit level: why greater equality makes societies stronger. East-West Publishing House, S., 2014 (2010). (in Bulgarian)

[20] http://nsi.bg/bg/content/2994/              -  -            -  - - -                  (Population forecast by age and gender – in Bulgarian)

3.2. There is a direct link between **social minimum and paying taxes**. Assuming that the poverty threshold in Bulgaria in 2016 was BGN 300 per month, this means that a family with 3 children (under 18 years of age or continuing their studies) must have a minimum of BGN 1,500 per month. In other words, if the two parents earn a total income of BGN 1,200, then their income tax will be BGN -300 (Negative Income Tax[21]). "The negative income tax" will eliminate all complementary social benefits – for heating, lighting, assistance, etc. Conversely, if another family with the same number of children has a monthly income (the same monthly amount throughout the whole year!) in the amount of BGN 8,000, then the tax will be levied after the deduction of BGN 1,500. Of course, there are other options.

3.3. The most difficult task will be to **"educate" the public and find enough "fans" of the progressive tax rate.** It is recommended that the tax brackets be introduced annually in order to prepare the public. For example, the lowest tax rate could be 0% for household members earning less than BGN 300 per month; about 10% for all household members earning between BGN 300-1,000 per month; 12% for incomes between BGN 1,001-2,000; 14% for incomes between BGN 2,001-4,000; 16% for incomes between BGN 4,001-6,000, 18% for incomes between BGN 6,001-8,000, 20% for incomes between BGN 8,001-10,000. "Work more and more innovatively, earn more, consume more!". The tax bracket scale is provisional, but after the Ministry of Finance makes additional calculations, it can take a more specific form.

The fine-tuning of the tax system must ensure not only the implementation of new policies but also the achievement of a great public consensus.

## Conclusions

Although justice, like freedom, is an element and manifestation of the "imagined reality" in citizens' minds, the state must comply with it. In particular, **the tax system has to adapt to the mass imagined reality in order for it to appear fairer and more acceptable**.

## Reference literature

*Angelov, Ivan*, The truth about the low "flat" tax rates.
http://epicenter.bg/article/Istinata-za-niskite--ploski--danatsi/115218/11/33 (in Bulgarian)

---

[21] An income maintenance scheme whereby additional funds are to be paid to those citizens whose income is below the threshold set by the government (this is usually the calculated subsistence minimum). The advocates of this scheme claim that it can eliminate the poverty trap, reduce administrative costs, and increase the influx of workers from low-income households as they are no longer negatively affected by the high marginal tax rates if they switch from social benefits to income earned under an employment contract.